# *Evaluating Mental Stress Among College Students Using Heart Rate and Hand Acceleration Data Collected from Wearable Sensors*

Moein Razavi[1], Anthony McDonald[2], Ranjana Mehta[2], Farzan Sasangohar[1*]

[1] Wm Michael Barnes Department of Industrial and Systems Engineering, Texas A&M university

[2] Department of Industrial and Systems Engineering, University of Wisconsin-Madison

*Corresponding author: sasangohar@tamu.edu; 979-458-2337; 3131 TAMU, College Station, TX 77843, United States

**Abstract**

Stress is various mental health disorders including depression and anxiety among college students. Early stress diagnosis and intervention may lower the risk of developing mental illnesses. We examined a machine learning-based method for identification of stress using data collected in a naturalistic study utilizing self-reported stress as ground truth as well as physiological data such as heart rate and hand acceleration. The study involved 54 college students from a large campus who used wearable wrist-worn sensors and a mobile health (mHealth) application continuously for 40 days. The app gathered physiological data including heart rate and hand acceleration at one hertz frequency. The application also enabled users to self-report stress by tapping on the watch face, resulting in a time-stamped record of the self-reported stress. We created, evaluated, and analyzed machine learning algorithms for identifying stress episodes among college students using heart rate and accelerometer data. The XGBoost method was the most reliable model with an AUC of 0.64 and an accuracy of 84.5%. The standard deviation of hand acceleration, standard deviation of heart rate, and the minimum heart rate were the most important features for stress detection. This evidence may support the efficacy of identifying patterns in physiological reaction to stress using smartwatch sensors and may inform the design of future tools for real-time detection of stress.

*Keywords*: Mental health, Machine learning, Stress, Students, Detection

## 1. Introduction

Exposure to prolonged and chronic stress may increase the risk of impaired physical and mental health ("American Psychiatric Association," 2000; Tafet & Nemeroff, 2016). Certain demographics and populations are more prone to high stress and its associated mental health disorders. One such population is college students—a recent study conducted in 23 countries indicated that over 20% of college students suffer from at least one mental health disorder (Auerbach et al., 2016). Another study reported that stress-induced anxiety is the most common and severe health issue among college students (Gao et al., 2020). However, despite the alarming rate of mental stress among students, a majority of college students do not seek help and treatment (Bravo et al., 2018). Given the potential stigma associated with mental health, self-management regimens may provide a more effective care pathway. Such self-management care may require access to just-in-time digital therapeutics or healthcare providers. Stress awareness may play a key role in enabling or triggering timely interventions. Indeed, stress awareness through frequent self-reporting on mobile health applications followed by just-in-time therapeutic action have shown to facilitate stress awareness and management (Morris et al., 2010).

Another approach to support stress awareness is through automated stress detection. Most research in the area of stress detection utilizes physiological markers and advanced statistical and artificial intelligence tools such as machine learning to provide a momentary assessment of stress (Healey & Picard, 2005; Koelstra et al., 2012; Kyriakou et al., 2019; Šalkevicius et al., 2019; Wen et al., 2017). For instance, prior research used heart rate and wrist-worn accelerometers (McDonald et al., 2019; Sadeghi et al., 2022), skin conductance, skin temperature, hand acceleration (Umematsu et al., 2019), electroencephalography (Arsalan et al., 2019), heart rate variability and electrocardiography (Bichindaritz et al., 2018), along with self-reported surveys (Sadeghi et al., 2022), to predict individuals' stress level. Most commonly used machine learning techniques for stress detection included Neural Networks (NN), Support Vector Machines (SVM), Logistic Regression (LR) and Naïve Bayesian (NB), with high accuracies reported ranging from 80% to 94% using different biosignals mentioned above (see reviews done by Razavi et al., 2022, 2023). Giannakakis et al. (2022) provide comprehensive guidelines for the efficient detection of stress by investigating the impact of stress on multiple bodily responses and exploring reliable biosignal patterns and modeling methods (Giannakakis et al., 2022). Recent work has also examined the ecological validity of stress detection models in physically active contexts and in gender-specific populations using neural (functional near infrared spectroscopy) and heart rate variability signals (Karthikeyan et al., 2022). Similar studies in intensive care units have established that physiological variables such as heart rate and accelerometers may enable reliable, low-cost, and non-intrusive opportunities for automated stress detection (Ahmadi et al., 2022).

Despite the overall promise shown by the documented machine-learning-based stress detection efforts, the application of such methods in college student population remains a research gap. Wearable devices such as smart watches along with mobile health (mHealth) platforms, provide an opportunity for momentary assessment and just-in-time management of stress with practical advantages including affordability, non-intrusiveness, non-invasiveness, ease of use and versatility. In particular, wrist-worn (PPG)-based heart rate sensors have been used in the bulk of research in this area (McDonald et al., 2019; Nath et al., 2020; Sadeghi et al., 2022). However, despite the preliminary evidence of efficacy, practicality of stress detection using digital wearable platforms needs to be assessed in more naturalistic and longitudinal studies. In addition, machine learning approaches for stress detection lack training and testing split at the participant level which may impact the reliability and generalizability of the results (Betti et al., 2018; Vanman et al., 1998) due to individual differences in physiological markers and stress triggers (Varoquaux et al., 2017). Finally, the limited interpretation of the machine learning models may hinder the understanding of the factors that affect stress levels in individuals (Gjoreski et al., 2017; V & P, 2016).

To address these gaps, this research aimed at investigating a machine learning-based stress detection tool using physiological measures in a naturalistic study involving college students. This paper documents the study, the process of developing the detection algorithms, and interpretation of the results using SHapley Additive exPlanations (SHAP) values.

## 2. Materials and Methods

A longitudinal home study was conducted with college students in a large university in Southern United States. Students' self-reported and detected stress data was used to train a machine learning tool for stress detection.

## 2.1 Dataset

### 2.1.1 Participants

Fifty-four (54) college students (45 females, 9 males) were recruited from a large R1 university in Southern United States to participate in the study. Participants' age ranged from 18 to 34 (M = 21.65, SD = 3.5). Participants were recruited through university bulk email and psychological counselling services website. The criteria to participate in the study was being over 18 years old, owning an iPhone, having an active student status, and having sought help from the on-campus mental health support services. Those who showed interest to participate in the study were asked to complete a Perceived Stress Scale (PSS-10) and a Generalized Anxiety Disorder survey (GAD-7) and only participants who scored above 14 in PSS-10 (indicative of high stress) and 7 in GAD-7 (probabilistically indicative of Generalized Anxiety Disorder and at least moderate anxiety symptoms) were eligible for the study. Students received a $150 gift card for completing the study. This study was approved by the university's Institutional Review Board (Protocol ID: IRB2020-0162DCR).

### 2.1.2 Data Collection Procedure

After providing informed consent, participants received an Apple Watch Series 4 or 5. Participants then attended an onboarding session to learn about the study and data collection protocols and were instructed on how to pair and operate the smart watch and the mobile applications used for the data collection called "mental Health Evaluation and Lookout Program for college students (mHELP)" for iPhone and Apple Watch. The mHELP application was designed to gather physiological data, including heart rate and hand acceleration, at a frequency of 1 Hz. The application was set to run continuously in the background on the participants' Apple Watch, thus ensuring unobtrusive data collection.

The mHELP application featured a functionality that allowed users to self-report stress events both on the watch and phone apps (Fig. 1). These self-reported events were timestamped to create a record of stress responses. Participants were asked to only report instances of high stress which accompanied signs and symptoms such as increased heart rate, shallow breathing, muscle tension, or changes in behavior, such as difficulty concentrating or restlessness. The complete dataset, therefore, included three measures: heart rate, hand acceleration, and self-reported stress events.

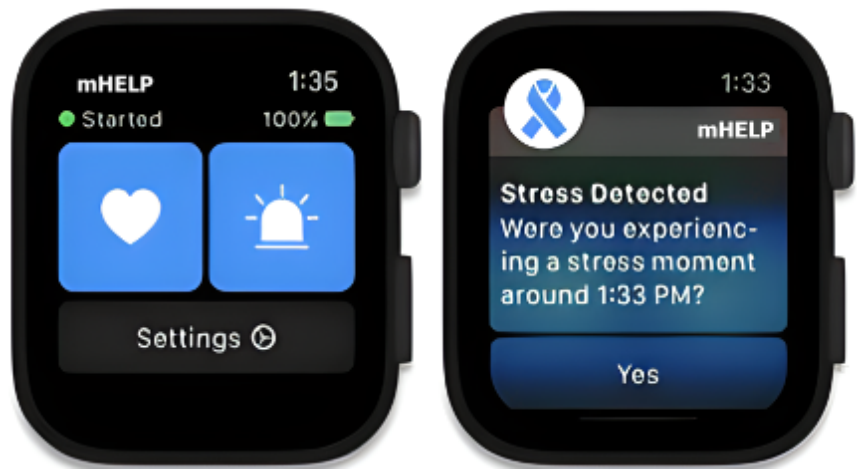

Fig. 1. mHELP Application

Participants were instructed to wear their watches continuously for a total of 40 days, except when charging was necessary, to ensure the maximum possible data collection. Non-identifiable data was automatically uploaded to an Amazon Web Services (AWS) S3 cloud server daily. A team of student researchers were tasked to monitor the data daily and contact the participants if there were any issues such as missing data.

## *2.2 Data Preprocessing*

Python 3.8.2 was used to conduct data preprocessing and to develop the machine learning algorithms. The data preprocessing included three steps: windowing and labeling, splitting the data into training and testing, and resampling the training dataset.

## 2.2.1 Windowing and Labeling

In line with previous research (Sadeghi et al., 2022), a 60-second window was analyzed for each stress episode (30 seconds before and 30 seconds after stress events were reported). The rest of the data after excluding stress events was labeled as non-stress events. The final dataset included 3,497 instances of stress events and 29,475 instances of non-stress events.

## 2.2.2 Training, Testing, and Upsampling

Since the number of events varied among participants, we divided the participants into training and test sets to achieve the closest ratio of train/test sets to a 80/20 split ratio. The training set included 34 participants (with 25608 events, 82.3% of the total data) and test set included 25 participants (with 5495 events, 17.7% of the total data). The data were unbalanced—92.4% of the events were non-stress events and 7% of the events were stress events. However, unbalanced datasets are common in clinical data (Arbain & Balakrishnan, 2019; Priya et al., 2020). To address this, we upsampled the training data to minimize the amount of information lost during the quantification process, to reduce noise, and improve resolution of results (Benchekroun et al., 2023; Salma & Ann, 2021). For upsampling, a ratio of 10 (non-stress events) to 7 (stress events) windows was chosen based on a sensitivity analysis. The number of stress and non-stress events are shown in Table 1.

Table 1. THE COUNT OF STRESS AND NON-STRESS EVENTS IN THE TRAINING AND TESTING SETS BEFORE AND AFTER RESAMPLING

|  | Label | Train Set | Test Set |
|---|---|---|---|
| **Before Resampling** | Non-Stress | 23668 (34 participants) | 5051 (25 participants) |
|  | Stress | 1940 (34 participants) | 444 (25 participants) |
| **After Resampling** | Non-Stress | 23668 | 5051 |
|  | Stress | 16567 | 444 |

## 2.2.3 Feature Generation and Selection

Features were generated separately for heart rate and acceleration data. The heart rate features included maximum heart rate, minimum heart rate, heart rate standard deviation, heart rate range, and mean heart rate over each window. These features were selected as they have been previously linked with stress (Bevilacqua et al., 2018; Sadeghi, McDonald, et al., 2021; Giannakakis et al., 2019). Acceleration data was first processed to calculate the absolute magnitude of hand acceleration (See Equation 1):

$$acceleration = \sqrt{a_x^2 + a_y^2 + a_z^2} \qquad (1)$$

where $a_x$ denotes hand acceleration in the $X$ direction, $a_y$ denotes hand acceleration in the $Y$ direction, and $a_z$ denotes hand acceleration in the $Z$ direction. Following this calculation time domain features including average hand acceleration, maximum hand acceleration, minimum hand acceleration, and range of hand acceleration were calculated for each window.

## 2.3. Model Assessment

To classify stress and non-stress events, six most prevalent machine learning models were trained: Random Forest, eXtreme Gradient Boosting (XGBoost), Generalized Linear Model (GLM), Linear Discriminant Analysis (LDA), Support Vector Machines (SVM) with Radial Basis Function (RBF) kernel, and K-Nearest Neighbors (KNN). The Area Under the Receiver Operating Characteristic (ROC) Curve (AUC) was used to evaluate the algorithms. Further, a 5x2 cross-validation test was applied to compare different models statistically (Dietterich, 1998). We applied hyperparameter tuning to the best-performed algorithm by performing a gridsearch on 240 combinations of parameters to find the combination with the best test accuracy. Table 2 shows the hyperparameters used for tuning.

Table 2. Hyperparameters used for tuning the best model (XGBoost)

| Hyperparameter | Value |
|---|---|
| Loss Function | Deviance/Exponential |
| Loss Criterion | Friedman Mean Squared Error (MSE)/MSE |
| Number of Estimators | 100/200 |
| Maximum Tree Depth | 3/5/7 |

## *2.4 Feature Importance and Model Interpretation*

To further investigate the features that contributed to model predictions, we used the SHAP method to calculate feature importance and correlations between features and predictions. The SHAP method allocates values to features in a model depending on their importance in prediction using game theoretic concepts. SHAP value summary plots were used to evaluate feature importance ranking as well as the distribution of each feature. The SHAP method has several advantages including being computationally efficient and matching human understanding (Lundberg & Lee, 2017).

## 3. Results

### *3.1 Model Performance*

As shown in Table 3 and Fig. 2, XGBoost achieved the highest AUC. A pairwise t-test comparing the results obtained from 5x2 cross-validation accuracies showed that XGBoost significantly outperformed SVM (t = 5.48, $p$ = 0.01) and GLM (t = 2.53, $p$ = 0.04). XGBoost did not perform significantly better than Random Forest (t = 0.66, $p$ = 0.53), LDA (t = 1.16, $p$ = 0.3), and KNN (t = 0.43, p = 0.68). Further, according to evaluation results (shown in Table 3), XGBoost performed the best in terms of recognizing stress events (true positives). Table 4 shows the confusion matrix for each model at different cut-off thresholds. The best combination obtained for hyperparameters based on test accuracy is shown in Table 5. The test accuracy and AUC of this model were 84.5% and 0.57, respectively.

Table 3. TRAINING AND VALIDATION RESULTS FOR THE TRAINED MACHINE LEARNING ALGORITHMS

|  | Train Accuracy | Test Accuracy | 10-fold CV Score | AUC |
|---|---|---|---|---|
| **Random Forest** | 99.7 | 91.0 | 98.6 | 0.59 |
| **XGBoost** | 69.6 | 74.0 | 67.6 | 0.66 |
| **GLM** | 61.2 | 76.3 | 61.5 | 0.65 |
| **LDA** | 62.2 | 76.4 | 61.5 | 0.65 |
| **SVM** | 67.6 | 75.8 | 65.9 | 0.62 |
| **KNN** | 92.8 | 73.0 | 86.5 | 0.55 |

Abbreviations: XGBoost = eXtreme Gradient Boosting; LDA = Linear Discriminant Analysis; SVM = Support Vector Machines; KNN = K-Nearest Neighbors, CV = Cross Validation

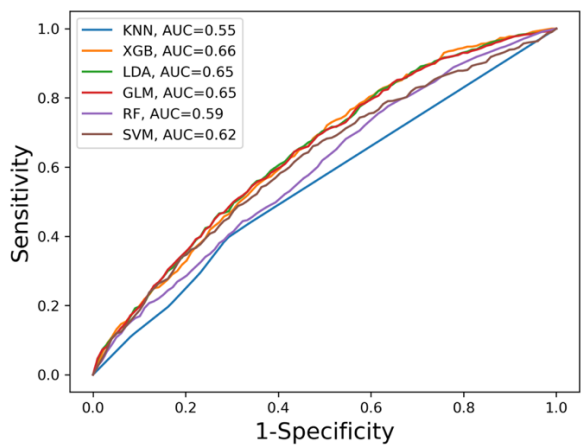

Fig. 2. AUROC for different trained machine learning algorithms. Abbreviations: KNN = K-Nearest Neighbors; XGBoost = eXtreme Gradient Boosting; LDA = Linear Discriminant Analysis; GLM = Generalized Linear Model; RF = Random Forest; SVM = Support Vector Machines.

Table 4. CONFUSION MATRICES FOR TRAINED MACHINE LEARNING ALGORITHMS AT DIFFERENT PROBABILITY CUT-OFF THRESHOLDS

| Design Scenario | Algorithm | TP | FN | FP | TN | TPR | FPR |
|---|---|---|---|---|---|---|---|
| Prioritize stress event detection (TPR[a]=1) | RF | 444 | 0 | 5051 | 0 | 1 | 1 |
|  | XGBoost | 444 | 0 | 4999 | 52 | 1 | 0.99 |
|  | GLM | 444 | 0 | 5051 | 0 | 1 | 1 |
|  | LDA | 444 | 0 | 5051 | 0 | 1 | 1 |
|  | SVM | 444 | 0 | 5051 | 0 | 1 | 1 |
|  | KNN | 444 | 0 | 4899 | 152 | 1 | 0.97 |
| Balanced priorities | RF | 223 | 221 | 0 | 5051 | 0.5 | 0 |
| (TPR=0.5) | XGBoost | 224 | 220 | 404 | 4647 | 0.5 | 0.08 |
| | GLM | 222 | 222 | 1566 | 3485 | 0.5 | 0.31 |
| | LDA | 222 | 222 | 1515 | 3536 | 0.5 | 0.30 |
| | SVM | 222 | 222 | 1616 | 3435 | 0.5 | 0.32 |
| | KNN | 222 | 222 | 808 | 4243 | 0.5 | 0.16 |
| Prioritize false positive minimization (FPR[b]=0.1) | RF | 231 | 213 | 504 | 4547 | 0.52 | 0.1 |
| | XGBoost | 293 | 151 | 503 | 4548 | 0.66 | 0.1 |
| | GLM | 93 | 351 | 508 | 4545 | 0.21 | 0.1 |
| | LDA | 89 | 355 | 505 | 4546 | 0.2 | 0.1 |
| | SVM | 155 | 289 | 505 | 4546 | 0.35 | 0.1 |
| | KNN | 133 | 311 | 498 | 4553 | 0.3 | 0.1 |

Abbreviations: XGBoost = eXtreme Gradient Boosting; LDA = Linear Discriminant Analysis; SVM = Support Vector Machines; KNN = K-Nearest Neighbors, RF = Random Forest, GLM = Generalized Linear Model

[a] True Positive Rate; [b] False Positive Rate

Table 5. BEST HYPERPARAMETERS OBTAINED BY GRIDSEARCH

| Hyperparameter | Value |
|---|---|
| Loss Function | Deviance |
| Loss Criterion | MSE |
| Number of Estimators | 20 |
| Maximum Tree Depth | 7 |

## 4. Model interpretation

We used SHAP values to interpret our best machine learning model, XGBoost. Using SHAP values we obtained the most important features in prediction of stress using XGBoost model. We also found how (in what direction) each parameter affects stress (whether increasing the value of each parameter shows an increasing likelihood of detecting a stress event or non-stress event). The SHAP summary plot for the XGBoost model is shown in Fig. 3.

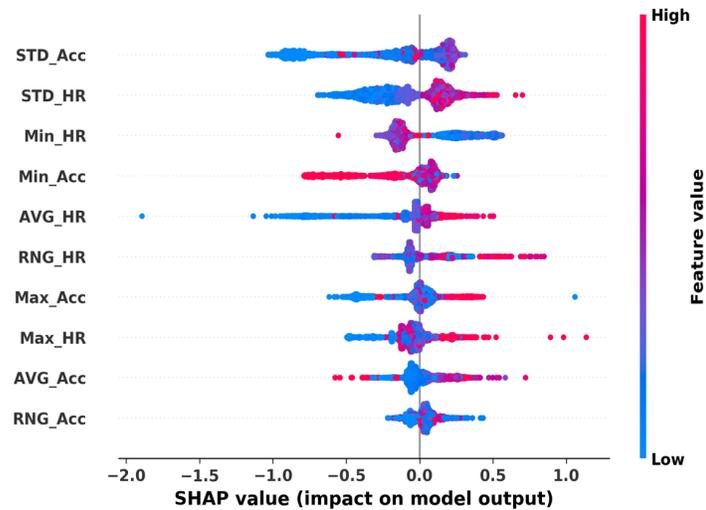

Fig. 3. SHAP value summary plot for all features of XGBoost model. Y-axis shows the features importance list based on features effect on the prediction. The X-axis indicates log-odds of stress event prediction and displays how each variable influences the model's output. The color in the figure represents the relevance of the variables' value in predicting the output. The most relevant hand acceleration characteristic is the standard deviation of this variable (STD_Acc), and the most significant heart rate time-domain variables for predicting stress events are the minimum heart rate (Min_HR) and heart rate standard deviation (STD_HR).

The SHAP dependence plots for the three most significant variables are shown in Fig. 4. SHAP dependency plots indicate the contribution of a given variable to the XGBoost model considering the variables' distribution.

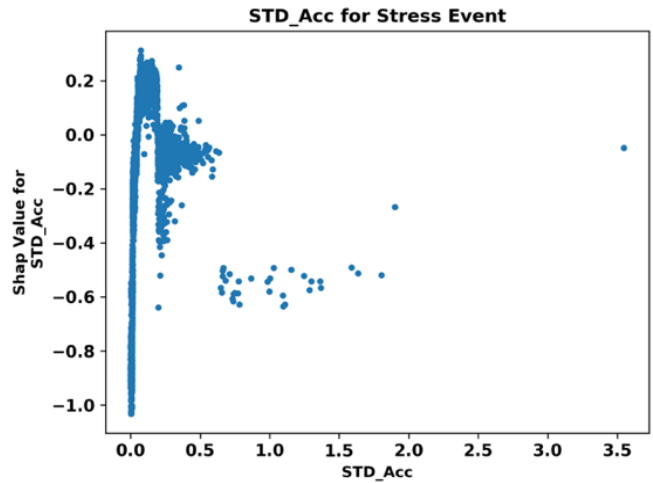

(a)

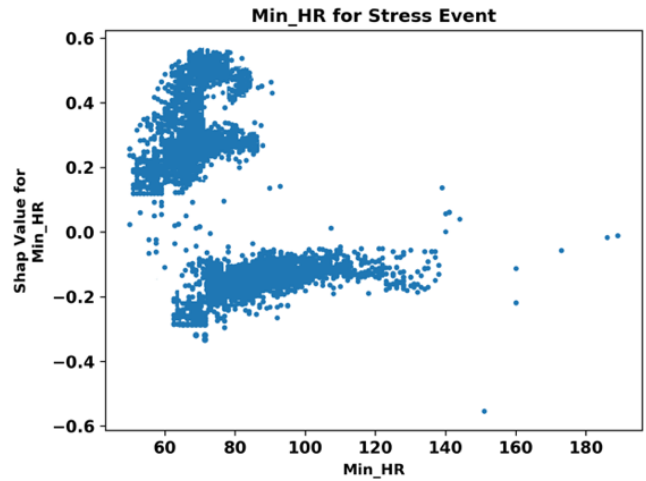

(b)

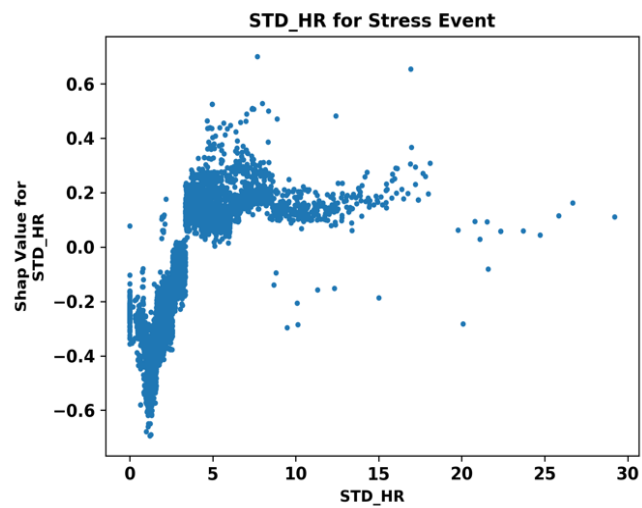

(c)

Fig. 4. SHAP dependence plots, a) SHAP plot for the standard deviation of hand acceleration (m/s$^2$), b) SHAP plot for minimum heart rate (bpm), c) SHAP plot for heart rate standard deviation (bpm). Each point corresponds to each observation from the dataset: X value represents the variable value for that observation, and Y value reflects the SHAP value, representing the influence of that variable (with that given value on the X-axis) on the prediction. The unit if the X-axis is the same as the variable unit, and the unit for Y-axis is log-odds of predicting a stress event.

As illustrated in Fig. 4, when the standard deviation of heart rate increases, the likelihood of predicting stress events increases meaning that higher heart rate fluctuations are associated with higher chances of perceiving stress. Moreover, stress events are more likely to be detected with higher minimum heart rate values across the windows. When the minimum heart rate increased, in most cases SHAP values increased as well. Hence, higher minimum heart rate values have more significant impact on detecting stress events. Finally, hand acceleration standard deviation has a nonlinear relationship effect on predicting the output.

## 5. Discussion

Using heart rate and accelerometer data collected from non-invasive off-the-shelf wearable sensors, we developed, assessed, and interpreted multiple machine learning algorithms to detect stress events in college students. Subjectively reported stress events collected in a naturalistic study served as the ground truth. Of the six algorithms trained in this study, the XGBoost method was the most robust, with an AUC of 0.64 and an 84.5% accuracy. The SHAP summary plot showed that the top hand acceleration feature was standard deviation of hand acceleration, and two top time-domain heart rate features were heart rate standard deviation and minimum heart rate.

Further analysis using the SHAP dependency plots demonstrated that the hand acceleration standard deviation exhibits non-linear correlations with the detection of stress events. Similar relationship has been reported for individuals diagnosed with PTSD (Sadeghi, McDonald, et al., 2021). These findings support the previous evidence (e.g., Garcia-Ceja et al., 2016; Sano & Picard, 2013; Wong et al., 2019) connecting physical activity with mental health outcomes and in particular stress. This may suggest the need for inclusion of physical activity in care pathways for students with mental health concerns.

Next, according to the SHAP dependency plots, heart rate standard deviation and minimum heart rate were the most important features influencing the algorithm's ability to detect stress events. More specifically, the odds of detecting stress events increase as the heart rate standard deviation increases, i.e., as heart rate fluctuates more. This finding is in line with prior research that indicated individuals' heart rate fluctuate more significantly during high stress and anxiety events (Sadeghi, McDonald, et al., 2021; Sadeghi, Sasangohar, et al., 2021). Furthermore, the results showed that there is a non-linear relationship between stress events and minimum heart rate, which may suggest a higher sensitivity to perceive stress at lower ranges of heart rate during a stress event.

This study offers several unique contributions compared to prior research on stress detection (Asha et al., 2021; Melillo et al., 2011; Salahuddin & Kim, 2006). First, we established the efficacy of utilizing sensors provided on a smartwatch for a machine learning-enabled stress detection mechanism. While more sophisticated heart rate and other physiological sensors with higher sampling rates may improve the accuracy of detection, optical heart rate measurements are increasingly prevalent in real-world settings due to their non-invasive properties, making them a practical choice for widespread stress monitoring (Building & Auditorium, 2018). Also, to the best of our knowledge, this is the first study that uses perceived stress data in naturalistic settings which may provide a more ecologically valid assessment of the phenomenon of interest. While induced stress in a lab setting provides an opportunity for a controlled observation of effects, such stress may not be representative of real-world stress. Finally, to our knowledge this is the first study investigating real-time automated monitoring of stress among college students.

*Limitations and Future Work*

This study had several noteworthy limitations. First, the study involved students from one university campus in the United States and the findings may not be generalizable to college students in other universities or countries. Second, we used a relatively low sampling rate of 1 Hz for acceleration. While this was necessary to optimize the battery life and reduce the attrition associated with the discomfort of charging the smart watch multiple times a day, more work is needed to evaluate the findings using higher grade acceleration sensors and higher sampling rate. Third, individual differences in stress perception and reporting might have biased the results. Future work may address such variability in reporting by implementing more nuanced operationalization of a subset of stress events with distinct signs and symptoms. Fourth, our reliance on noninvasive wearable devices, while innovative, comes with its own set of challenges. These devices often have restrictions like limited precision and constrained time windows for data capture. Also, conducting the study in a naturalistic environment, while providing a more ecologically valid assessment, introduces potential sources of noise and data loss. Fifth, similarity between individuals' physiological data in the training and testing sets is challenging. Although we addressed this by segregating testing and training datasets at the participant level, more work is warranted to address this limitation. Sixth, mainly due to limited sample, the full potential of big data algorithms, such as neural networks, was not explored in this study, suggesting areas for further refinement in future research. Seventh, the relatively high false alarm in our detection method might affect its acceptance and reporting among students. Future work should assess students' acceptance thresholds for such false alarm rates. Finally, external factors, such as physical activity or other stimuli, could influence heart rate values, emphasizing the intricate nature of stress detection and the importance of considering multiple physiological parameters.

## 6. Conclusion

College students' mental health concerns are on the rise and there is a timely need to investigate effective care pathways for this vulnerable population. Advances in artificial intelligence and mobile health provide an opportunity to support non-intrusive monitoring of mental health states, enable self-management, and improve the quality of care. This study evaluated the efficacy of building a machine learning tool that can learn from students' self-report of stress events in naturalistic settings to detect physiological patterns in response to stress. The findings presented here suggest that both heart rate and acceleration features may be used to detect such patterns for stress. Future work is needed to assess utilizing such tools for real-time detection of stress among college students.

18Dietterich, T. G. (1998). Approximate Statistical Tests for Comparing Supervised Classification Learning Algorithms. *Neural Computation*, *10*(7), 1895–1923. https://doi.org/10.1162/089976698300017197

Durán Acevedo, C. M., Carrillo Gómez, J. K., & Albarracín Rojas, C. A. (2021). Academic stress detection on university students during COVID-19 outbreak by using an electronic nose and the galvanic skin response. *Biomedical Signal Processing and Control*, *68*, 102756. https://doi.org/10.1016/j.bspc.2021.102756

Gao, W., Ping, S., & Liu, X. (2020). Gender differences in depression, anxiety, and stress among college students: A longitudinal study from China. *Journal of Affective Disorders*, *263*, 292–300. https://doi.org/10.1016/J.JAD.2019.11.121

Garcia-Ceja, E., Osmani, V., & Mayora, O. (2016). Automatic Stress Detection in Working Environments From Smartphones' Accelerometer Data: A First Step. *IEEE Journal of Biomedical and Health Informatics*, *20*(4), 1053–1060. https://doi.org/10.1109/JBHI.2015.2446195

Giannakakis, G., Grigoriadis, D., Giannakaki, K., Simantiraki, O., Roniotis, A., & Tsiknakis, M. (2019). Review on psychological stress detection using biosignals. *IEEE Transactions on Affective Computing*.

Giannakakis, G., Grigoriadis, D., Giannakaki, K., Simantiraki, O., Roniotis, A., & Tsiknakis, M. (2022). Review on Psychological Stress Detection Using Biosignals. *IEEE Transactions on Affective Computing*, *13*(1), 440–460. https://doi.org/10.1109/TAFFC.2019.2927337

Gjoreski, M., Luštrek, M., Gams, M., & Gjoreski, H. (2017). Monitoring stress with a wrist device using context. *Journal of Biomedical Informatics*, *73*, 159–170. https://doi.org/10.1016/J.JBI.2017.08.006

Healey, J. A., & Picard, R. W. (2005). Detecting stress during real-world driving tasks using physiological sensors. *IEEE Transactions on Intelligent Transportation Systems*, *6*(2), 156–166. https://doi.org/10.1109/TITS.2005.848368

21Salma, M. U., & Ann, K. A. A. (2021). Active Learning from an Imbalanced Dataset: A Study Conducted on the Depression, Anxiety, and Stress Dataset. In *Handbook of Machine Learning for Computational Optimization*. CRC Press.

Sano, A., & Picard, R. W. (2013). Stress Recognition Using Wearable Sensors and Mobile Phones. *2013 Humaine Association Conference on Affective Computing and Intelligent Interaction*, 671–676. https://doi.org/10.1109/ACII.2013.117

Schubert, C., Lambertz, M., Nelesen, R. A., Bardwell, W., Choi, J. B., & Dimsdale, J. E. (2009). Effects of stress on heart rate complexity—A comparison between short-term and chronic stress. *Biological Psychology*, *80*(3), 325. https://doi.org/10.1016/J.BIOPSYCHO.2008.11.005

Tafet, G. E., & Nemeroff, C. B. (2016). The Links Between Stress and Depression: Psychoneuroendocrinological, Genetic, and Environmental Interactions. *J Neuropsychiatry Clin Neurosci*, *28*, 77–88. https://doi.org/10.1176/appi.neuropsych.15030053

Umematsu, T., Sano, A., & Picard, R. W. (2019). *Daytime Data and LSTM can Forecast Tomorrow's Stress, Health, and Happiness*. Proceedings of the Annual International Conference of the IEEE Engineering in Medicine and Biology Society, EMBS. https://doi.org/10.1109/EMBC.2019.8856862

V, V., & P, K. (2016). Real time stress detection system based on EEG signals. *Biomedical Research*, *0*(0), 271–275.

Vahdani, M., Yaraghi, S., Neshasteh, H., & Shahabadi, M. (2019). Narrow-Band 4.3μm Plasmonic Schottky-Barrier Photodetector for $CO_2$ Sensing. *IEEE Sensors Letters*, *3*(3), 1–4. https://doi.org/10.1109/LSENS.2019.2895968

Vanman, E. J., Dawson, M. E., & Brennan, P. A. (1998). Affective reactions in the blink of an eye: Individual differences in subjective experience and physiological responses to emotional stimuli. *Personality and Social Psychology Bulletin*, *24*(9), 994–1005. https://doi.org/10.1177/0146167298249007